\documentclass[preprint,superscriptaddress,nofootinbib,amsmath,amssymb,aps]{revtex4}

\usepackage{graphicx}% Include figure files
\usepackage{dcolumn}% Align table columns on decimal point
\usepackage{bm}% bold math
\usepackage{hyperref}% add hypertext capabilities
\usepackage{xspace,xcolor}
\usepackage{float}
\usepackage{multirow}
\usepackage{xspace}
\usepackage{physics,ulem,slashed}

\newcommand{\gev}{\operatorname{GeV}}
\newcommand{\tev}{\operatorname{TeV}}
\newcommand{\ab}{\operatorname{ab}}

\newcommand{\met}{\ensuremath{\not\mathrel{E}_T}\xspace}

\newcommand{\emc}{$e\mu$ {\text collider}\xspace}
\newcommand{\eec}{$e^+ e^-$ {\text collider}\xspace}
\newcommand{\mmc}{$\mu^+ \mu^-${\text collider}\xspace}

\newcommand{\mmcs}{$\mu^+ \mu^-$ {\text colliders}\xspace}
\newcommand{\kh}{\ensuremath{\kappa_{h}}\xspace}
\newcommand{\khhww}{\ensuremath{\kappa_{hhWW}}\xspace}
\newcommand{\khww}{\ensuremath{\kappa_{hWW}}\xspace}
\newcommand{\khhvv}{\ensuremath{\kappa_{hhVV}}\xspace}
\newcommand{\khvv}{\ensuremath{\kappa_{hVV}}\xspace}

\newcommand{\pythia}{\textsc{Pythia8}\xspace}
\newcommand{\delphes}{\textsc{Delphes}\xspace}
\newcommand{\fastjet}{\textsc{Fastjet}~\cite{fastjet}\xspace}
\newcommand{\sqrts}{\sqrt{s}}

\allowdisplaybreaks[4]

\makeatletter

\begin{document}

\title{Measuring $hhWW$ Coupling at Lepton Colliders}

\author{Qing-Hong Cao}
\email{qinghongcao@pku.edu.cn}
\affiliation{School of Physics, Peking University, Beijing 100871, China}
\affiliation{Center for High Energy Physics, Peking University, Beijing 100871, China}

\author{Kun Cheng}
\email{chengkun@pku.edu.cn}
\affiliation{School of Physics, Peking University, Beijing 100871, China}

\author{Yandong Liu}
\email{ydliu@bnu.edu.cn}
\affiliation{Key Laboratory of Beam Technology of Ministry of Education, College of Nuclear Science and Technology, Beijing Normal University, Beijing 100875, China}
\affiliation{Beijing Radiation Center, Beijing 100875, China}

\author{Xiao-Rui Wang}
\email{xiaorui\_wong@pku.edu.cn}
\affiliation{School of Physics, Peking University, Beijing 100871, China}

\begin{abstract}
The quartic Higgs-gauge-boson coupling $g_{hhWW}$ is sensitive to the electroweak symmetry breaking
mechanism, however, it is challenging to be measured at the large hadron collider. We show that the coupling can be well probed at future lepton colliders through di-Higgs boson production via the vector boson fusion channel.   We perform a detailed simulation of $\ell^+\ell^-\to \nu\bar\nu hh\to 4b+\slashed{E}_T$ with parton showering effects at $e^+e^-$, $\mu^+\mu^-$ and $e^-\mu^+$ colliders. We find that the regions of $g_{hhWW}/g_{hhWW}^{\rm SM}<0.86$ and $g_{hhWW}/g_{hhWW}^{\rm SM}>1.32$ can be discovered at the $5\sigma$ confidence level at the 10 TeV \mmc with an integrated luminosity of 3 ab$^{-1}$.
\end{abstract}

\maketitle

\section{Introduction}
The discovery of the Higgs boson~\cite{ATLAS:2012yve,CMS:2012qbp} has significant implications for the study of its characteristics, including its mass, spin, and interactions with other particles in the Standard Model (SM). 
Yet, our understanding of the electroweak symmetry breaking (EWSB) mechanisms and the true character of the Higgs boson remains shrouded in mystery. The couplings of the Higgs boson to $W^\pm$ and $Z$ bosons, $g_{hVV}$ and $g_{hhVV}$, pivotal for insights into EWSB, are intrinsically linked to the interactions of the gauge interactions and pattern of EWSB. 
The SM values of the couplings, $g^{\text{SM}}_{hVV}$ and $g^{\text{SM}}_{hhVV}$, are uniquely fixed by the weak charge of the scalar doublet. The interrelationship of these couplings manifests the internal consistency of the SM and ensures perturbative unitarity of gauge boson scatterings.

These couplings, however, can be altered in new physics (NP) models. 
To describe the NP effects in these couplings, we define $\kappa_{hhVV}$ and $\kappa_{hVV}$ as:
\begin{align}
\kappa_{hhVV}\equiv \frac{g_{hhVV}}{g^{\text{SM}}_{hhVV}}, \quad \kappa_{hVV}\equiv \frac{g_{hVV}}{g^{\text{SM}}_{hVV}}.
\end{align}
Any deviation of $\kappa_{hhVV}$ and $\kappa_{hVV}$ from unity implies the existence of NP.  
For instance, a singlet scalar extension model predicts that both the $g_{hhVV}$ and $g_{hVV}$ couplings will decrease compared to the SM due to mixings, leading to $\kappa_{hhVV}$ and $\kappa_{hVV}$ that are less than one. 
Doublet scalar extension models might decrease $g_{hVV}$, but leave $g_{hhVV}$ unchanged due to EWSB constraints. Consequently, in these models, $\kappa_{hVV}$ is less than one, but $\kappa_{hhVV}$ retains the SM value. Furthermore, a larger $\kappa_{hVV/hhVV}$ is predicted in the scalar triplet or larger representations, owing to increasing Clebsch-Gordan (CG) coefficients. On the other hand, both the couplings $g_{hVV}$ and $g_{hhVV}$ are reduced in the Composite Higgs Model (CHM), in which the coupling ratio $\kappa_{hVV}=\sqrt{1-\xi}$ and $\kappa_{hhVV}=1-2\xi $,  where $\xi=v^2/f^2$, $v$ is the electroweak symmetry breaking scale and $f$ represents the compositeness scale~\cite{Agashe:2004rs}.

Measuring $\khhvv$ and $\khvv$ and investigating their relation would shed lights on the mechanism of EWSB. The ratio $\khvv$ can be precisely pinned down by the vector boson fusion (VBF) type single Higgs production at the LHC~\cite{CMS:2018uag,Sharma:2022epc} as $\kappa_{hVV}$ can be extracted directly from $\khvv = \sqrt{\sigma_h/\sigma_h^{\rm SM}}$, where $\sigma_h$ is the cross section of VBF-type single Higgs production and $\sigma_{h}^{\rm SM}$ is the SM prediction.
However, it is difficult to measure the ratio \khhvv at the LHC, because \khhvv can only be measured from VBF type di-Higgs production processes while the gluon fusion processes is the dominant di-Higgs production processes at the LHC~\cite{deFlorian:2013jea}.
Presently, the LHC provides a loose constraint on $\khhvv$ with  $0<\khhvv<2.1$~\cite{ATLAS:2020jgy,ATLAS:2023qzf}.

In our study, we investigate the VBF-type di-Higgs production at future high-energy lepton colliders~\cite{Roloff:2019crr,Han:2020pif}, and we focus on the measurement of Higgs-gauge-boson couplings.  The rest of this paper is organized as follows.  The cross section of different di-Higgs production is described in Sec.~\ref{sec:XS}. The selection and reconstruction of events is presented in Sec.~\ref{sec:simulation}, and the result is shown in Sec.~\ref{sec:analysis}. Finally we conclude in Sec.~\ref{sec:conclusion}.

\section{Di-Higgs production cross section at future lepton colliders}\label{sec:XS}

\begin{figure}
    \centering
    \includegraphics[width=.7\linewidth]{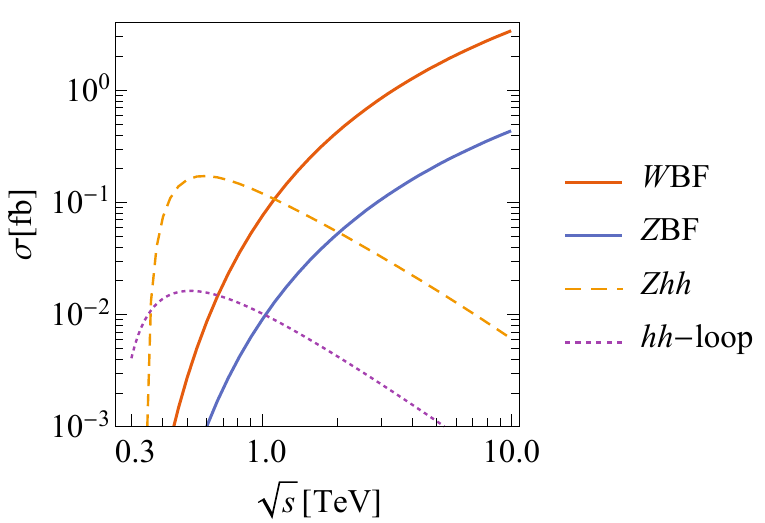}
    \caption{The cross sections of di-Higgs boson production at lepton colliders as a function of colliding energies ($\sqrt{s}$),
    where $W$BF and $Z$BF stands for the VBF process mediated by the $W$ and $Z$ boson, respectively.
    }
    \label{fig:Xsec}
\end{figure}

At high-energy lepton colliders, di-Higgs boson production can be instigated through several processes, including the VBF process, the $Z$-boson associated production ($Zhh$) process, and the loop-induced $hh$ pair production process. Figure~\ref{fig:Xsec} plots the cross section for each process as a function of the collision energy $\sqrts$. The contribution of loop-induced process is very small due to the loop suppression, and the $s$-channel production cross section of $Zhh$ decreases with collision energy when $\sqrts$ exceeds 1 TeV due to $s$-channel suppression. Contrarily, the VBF production cross section increases with $\log\sqrts/m_V$, as the the collinear splitting mechanism of the electroweak gauge bosons becomes the dominant phenomena of the initial state radiation at high energy~\cite{Han:2020uid}. Besides, the cross section of the charged current process (mediated by $W^\pm$) is approximately an order of magnitude larger than that of the neutral current process (mediated by $Z$). Therefore, we mainly focus on the $W^\pm$ fusion process in the subsequent analysis.

\begin{figure}
    \centering
    \includegraphics[width=\linewidth]{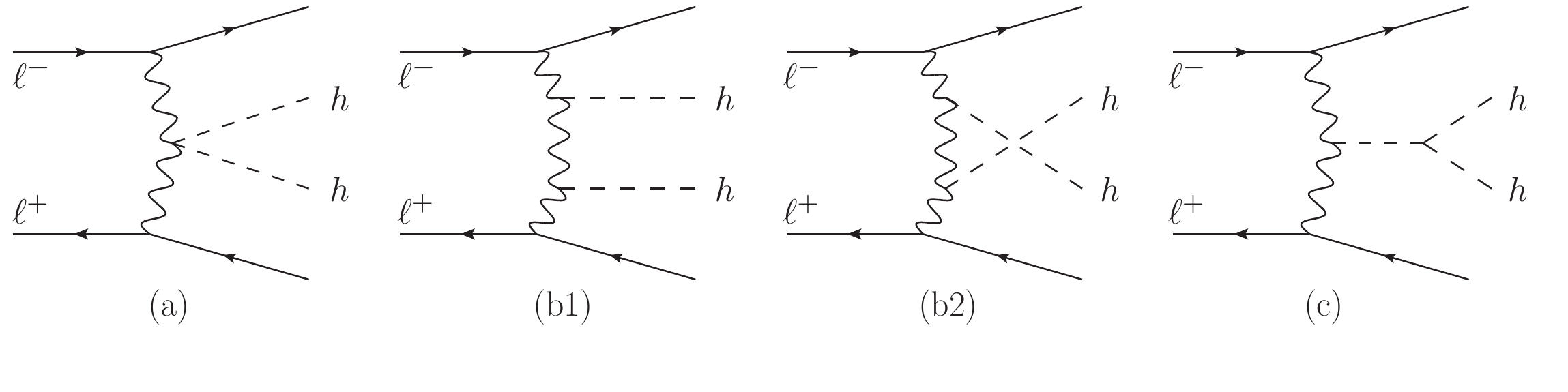}
    \caption{Feynman diagrams of the di-Higgs boson production via VBF processes at lepton colliders.}
    \label{fig:ll2HH}
\end{figure}

\begin{figure}
    \centering
    \includegraphics[scale=0.8]{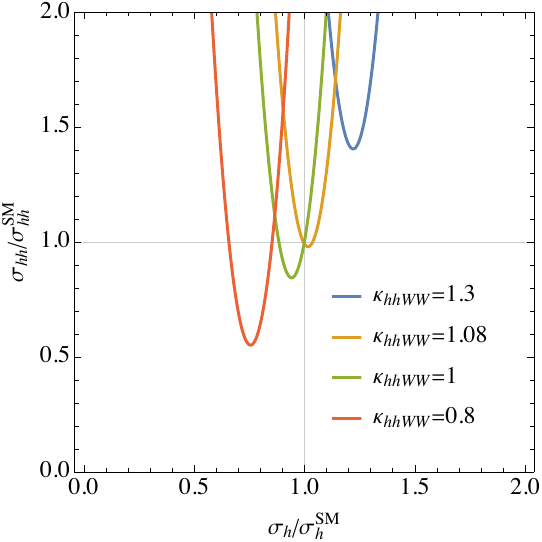}
    \caption{Contours of \khhww in the plane of $\sigma_{h}/\sigma_{h}^{\rm SM}$ and $\sigma_{hh}/\sigma_{hh}^{\rm SM}$, where $\sigma_h$ and $\sigma_{hh}$ presents the cross section of the VBF type single-Higgs boson production and the di-Higgs boson production at the 10 TeV muon collider, respectively. }
    \label{fig:correlation}
\end{figure}

\begin{figure}
    \centering
    \includegraphics[scale=0.62]{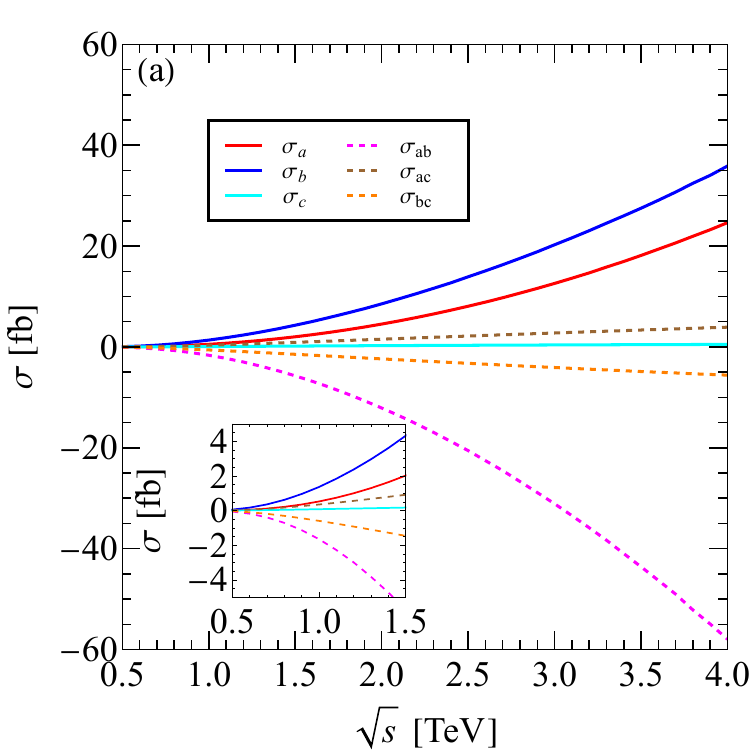}
    \caption{Cross section of the $W$-boson fusion process $\ell^+\ell^-\to \bar{\nu}_{\ell}\nu_{\ell}hh$ from the contribution of each diagram and their interferences. }
    \label{fig:cs}
\end{figure}

The VBF type di-Higgs production involves both the couplings $g_{hWW}$ and $g_{hhWW}$; see Fig.~\ref{fig:ll2HH}. Therefore, measuring $g_{hhWW}$ via di-Higgs production also relies on the value of $g_{hWW}$. Fortunately, the ratio \khww can be precisely measured from single Higgs production via VBF process and $Vh$ associated process. Combining both the single and di-Higgs production, one can determine the \khhww.
In Fig.~\ref{fig:correlation}, we plot the contours of \khhww in the plane of $\sigma_{h}/\sigma_{h}^{\rm SM}$ and $\sigma_{hh}/\sigma_{hh}^{\rm SM}$. Even though a degeneracy of \khhww appears for given total cross sections $\sigma_{h}$ and $\sigma_{hh}$, it can be resolved by analyzing the $m_{hh}$ distribution. In the work we performed simulations at three kinds of lepton colliders
\begin{itemize}
    \item \eec with collision energy of $\sqrts=1$ TeV;
    \item \emc with collision energies of $\sqrts=1$ TeV and $\sqrts=2$ TeV~\cite{Lu:2020dkx};
    \item \mmc with collision energy of $\sqrts=$10 TeV~\cite{Accettura:2023ked}.
\end{itemize}
Assuming the SM value of trilinear Higgs boson self-coupling, the coupling dependence of the cross section can be parametrized as
\begin{align}
    \sigma^{hh} =& \khhww^2 \sigma_a + \khww^4 \sigma_b + \khww^2 \sigma_c +\khhww\khww^2\sigma_{ab} +\khww^3\sigma_{bc}+\khhww\khww\sigma_{ac},
    \label{eq:sigmallvvhh1}
\end{align}
where $\sigma_a$, $\sigma_b$ and $\sigma_c$ parametrize the contribution from the Feynman diagram (a), (b) and (c) in Fig.~\ref{fig:ll2HH}, respectively; $\sigma_{ab}$, $\sigma_{ac}$ and $\sigma_{bc}$ denote the interference between each Feynman diagram.
For the process of $l^-l^+\to \nu_l \bar{\nu}_lhh$, the contribution from each Feynman diagram and their interference effects are shown in Fig.~\ref{fig:cs}, and we also show the values with the benchmark collision energies in Table~\ref{tab:cross_section1}.

\begin{table}
    \centering
    \caption{Parametrization of parton level cross sections of $l^-l^+\to\nu_l\bar{\nu}_l h h$ at lepton colliders.}
    \begin{tabular}{|c|c|c|c|c|c|c|}
    \hline
        cross section [fb] & $\sigma_{a} $ & $\sigma_{b} $ & $\sigma_{c} $ & $\sigma_{ab}$ &  $\sigma_{bc}$ &  $\sigma_{ac}$ \\
    \hline
        $1\tev$ \eec  & $0.54 $ & $1.41 $ & $0.09 $ &$-1.69 $& $-0.62 $ & $0.35 $\\
    \hline
        $1\tev$ \emc  & $0.54$ & $1.41$ & $ 0.09$ &$-1.69$& $-0.62$ & $0.35$\\
    \hline
        $2\tev$ \emc & $4.80$ & $9.02$ & $ 0.34$ &$-12.72$& $-2.53$ & $1.53$\\
    \hline
        $10\tev$ $\mu^+\mu^-$ collider   & 
         $177.2$ & $206.1$ & $ 0.88 $ &$-378.10$& $-12.11$ & $9.08$\\
    \hline
    \end{tabular}
    \label{tab:cross_section1}
\end{table}

When the ratio \khhww has a deviation from its SM value with $\delta\khhww$, namely $\khhww=1+\delta\khhww$, the leading contribution of the deviation arises from diagram (a) and its interference between other diagrams; see Eq.~\eqref{eq:sigmallvvhh1}.  It is observed in Fig.~\ref{fig:cs} that $\sigma_a$, $\sigma_b$ and $\sigma_{ab}$ increase with collision energy and dominate at high energy. Therefore, the cross section becomes more sensitive to \khhww with a larger collision energy.
The underlying physics of this explosion of $\sigma_a$, $\sigma_b$ and $\sigma_{ab}$ with respect to the energy is the violation of perturbative unitarity; see Appendix~\ref{app:unitarity} for details.  Therefore, the high energy lepton colliders are ideal to test NP models that change the Higgs-gauge-boson coupling $g_{hhWW}$.

\section{Collider Simulation}\label{sec:simulation}

We consider three lepton colliders, the \eec and \mmc with symmetric incident beam energies and the \emc with asymmetric incident beams, and we explore their potential of measuring the \khhww coupling through the VBF-type di-Higgs production.  Owing to the clean environment of the lepton colliders, one can reconstruct the two Higgs bosons in the $b\bar{b}$ decay mode, the mode with the largest branching ratio. This leads to a distinct signature comprising four $b$-jets and missing energy \met from two invisible neutrinos, in all the three lepton colliders.

The primary background processes for the $4b+$\met signal are $\ell_i^-\ell_j^+ \to \nu_i\bar{\nu}_j b\bar{b}b\bar{b}$ from on-shell $hZ/ZZ$ production, and $\ell^\pm \gamma \to b\bar{b}jj\nu\ell(\bar\nu\ell)$ from on-shell $hW^\pm/ZW^\pm$ production~\cite{Roloff:2019crr}. In addition, the $\ell^+\ell^-\to b\bar{b}b\bar{b}$ process, arising from on-shell $Zh$ or $ZZ$ production, may also contribute to the background at the \eec and \mmc, where \met is mimicked by undetected particles in the showering. Nevertheless, the reducible $4b$ backgrounds can be effectively suppressed by employing a relatively large \met cut or a recoil mass cut, as discussed below.

\subsection{Symmetric $e^+e^-$ and $\mu^+\mu^-$ Colliders}
First, we focus on the symmetric colliders, namely the $e^+ e^-$ and \mmcs.  
The \eec is proposed as a precision machine that benefits from the clear collision environment~\cite{ILC:2013jhg,Asner:2013psa,CEPCStudyGroup:2018ghi,ILCInternationalDevelopmentTeam:2022izu}.
The \mmc, as proposed in~\cite{Budker:1969cd,Parkhomchuk:1983ua,Neuffer:1983jr}, offers not only a clean environment but also a high collision energy~\cite{Palmer:1996gs,Stratakis:2022zsk,Abbott:2022jqq,Aime:2022flm,Han:2022ubw,Han:2022mzp}. While technical difficulties~\cite{Delahaye:2019omf,Long:2020wfp} arise from the short lifetime, they can be compensated by the advantage of increasing instantaneous luminosity with beam energy~\cite{Long:2020wfp}. We use \eec with $\sqrts=1$ TeV and \mmc with $\sqrts=10$ TeV as examples in our analysis.

In the following simulation, we generate parton-level events using MadEvent~\cite{madgraph5}, then pass them to \pythia~\cite{pythia} for showering and hadronization.  The showered events are clustered using \fastjet using anti-$k_T$ algorithm with clustering parameter $\Delta R=0.5$, and we subsequently simulate collider effects with \delphes~\cite{delphes}. We adopt the tracking and energy resolution modeling used in \delphes as specified in~\cite{CEPCStudyGroup:2018ghi}. We select a benchmark $b$-jet (mis-)tagging efficiencies from Ref.~\cite{CEPCStudyGroup:2018ghi} with the higher identification rate,
\begin{align}
p_{b\to b}=0.9,~~~p_{c\to b}=0.3,~~~p_{j\to b}=0.05,
\end{align}
where $j$ represents gluon or light quarks.

For the pre-selection, we require both the signal and backgrounds to satisfy the following cuts:
\begin{align}
    &n^\ell(p_T>10\gev)=0,~\met>10\gev,\nonumber \\
    &p_T^{jet}>15~\gev,~ -4.0<\eta^{jet}<4.0,~\Delta R^{mn} >0.5,
    \label{eq:precuts}
\end{align}
where $\ell$ denotes an electron or muon, $m (n)$ represents a reconstructed lepton or jet, and $\Delta R$ is the rapidity and azimuthal distance defined as $\Delta R^{mn}=\sqrt{ \Delta\eta_{mn}^2 + \Delta \phi_{mn}^2 }$. To eliminate backgrounds mediated by neutral current interactions, we require a \met signal and no isolated leptons with transverse momenta larger than $10\gev$.
To effectively reconstruct the two Higgs bosons, we require at least four $b$-jets and divide the four leading ones into two pairs by minimizing $\chi^2_{ijkl}=(m_{b_i b_j}-m_h)^2+(m_{b_k b_l}-m_h)^2$. The two pairs of $b$-jets are treated as two Higgs boson candidates. Then we impose the requirement that the invariant mass of the two $b$-jets in each pair must satisfy
\begin{equation}\label{eq:cutmbb}
100\gev<m_{bb}<150\gev,
\end{equation}
which enhances signal-to-backgrounds ratio dramatically; see Tables~\ref{tab:cutflowee1tev} and~\ref{tab:cutflow2}.

To further suppress electroweak resonance backgrounds, we require the likelihood of the four $b$-jets originating from the Higgs boson pair to be larger than that from a $ZZ$ pair or $Zh$ pair. We achieve this by employing the function $\chi^2(m_1,m_2)$, defined as
\begin{align}
\chi^2(m_1,m_2)\equiv \min_{{i,j,k,l}} \left[(m_{b_i b_j}-m_1)^2+(m_{b_k b_l}-m_2)^2 \right],
\end{align}
which represents the minimal mass difference between two $b$-jet pairs and $m_{1 (2)}$. We then require the four leading $b$-jets in each event to satisfy
\begin{align}\label{eq:cutresonant}
{\rm{HHCUT}:}~~~ \chi^2(m_h,m_h)< \chi^2(m_Z,m_Z),~~~\chi^2(m_h,m_h)< \chi^2(m_h,m_Z).
\end{align}
The selection criteria employed in this study effectively reduces the $ZZ$+\met and $Zh$+\met backgrounds and preserves the signal events, as illustrated in the fourth columns of Tables~\ref{tab:cutflowee1tev} and \ref{tab:cutflow2}.

At last, the process $e^+ e^-\to b\bar{b}b\bar{b}$ stemming from $Zh/ZZ$ resonant production often presents a non-negligible background. For example, at the $1\tev$ \eec, the cross section of these two non-intrinsic backgrounds are still larger than the signal cross section after imposing all the previous cuts. Although it is feasible to reject these backgrounds by increasing the $\met$ cut in Eq.~\eqref{eq:precuts}, this approach also reduces signal events considerably. A more effective strategy is to impose a cut on the recoil mass, defined as 
\begin{equation} \label{eq:recoilmass}
M_{\rm recoil}\equiv\sqrt{(p_1+p_2-p_{h_1}-p_{h_2})^2},
\end{equation}
where $p_1$ and $p_2$ represent the momenta of initial lepton, and $p_{h_1}$ and $p_{h_2}$ denote the momenta of reconstructed Higgs boson.  We require a hard cut on $M_{\rm recoil}$
\begin{align}
    M_{\rm recoil} \geq 200 \gev,
\end{align}
as the two neutrinos in the signal tend to fly back-to-back and lead to a large recoil mass. We find that, after the $M_{\rm recoil}$ cut, the non-intrinsic backgrounds are negligible, and the signal events almost remain the same; see Table~\ref{tab:cutflowee1tev}. With an integrated luminosity of $3\ab^{-1}$, the signal strength of the total cross section of di-Higgs production, $\mu_{hh}=\sigma_{hh}/\sigma_{hh}^{\rm SM}$, can be constrained to $0.35<\mu_{hh}<1.80$ at the $1\tev$ \eec and $0.9<\mu_{hh}<1.1$ at the $10\tev$ \mmc, at 95\% confidence level.

\begin{table}
    \centering
    \caption{Cut flow for the SM $\nu\bar{\nu} h h$  di-Higgs signal and backgrounds cross sections at $1\tev$ electron-positron collider (left)
    and 10 TeV muon collider (right).}
    \label{tab:cutflowee1tev}
    \begin{tabular}{|c|c|c|c|c|}
    \hline
        $\sigma$ [ab]& pre-cuts & $m_{bb}$ cut & HHCUT& $M_{\rm recoil}$ cut\\
    \hline
        Sig.  & 15.9 & 9.7 & 8.3 & 5.7\\
    \hline              
        $\nu\nu hZ$  & 54.4 & 11.1 & 6.3& 5.5\\
        $\nu\nu ZZ$  & 73.7 & 9.1 & 3.1& 2.9\\
        $\nu W^\pm h$  & 45.2 & 3.9 & 2.3 & 2.3 \\
        $\nu W^\pm Z$  & 47.0 & 2.5 & 1.1 & 1.1 \\
    \hline
        Bkg. & - & - & - & 11.8\\
    \hline
    \end{tabular}
    \begin{tabular}{|c|c|c|c|c|}
    \hline
        $\sigma$ [ab]  & pre-cuts & $m_{bb}$ cut & HHCUT & $M_{\rm recoil}$ cut\\
    \hline
        Sig. & 484.4 & 261.5 & 226.5 & 226.3  \\
    \hline              
        $\nu\nu hZ$  & 1163.1 & 168.6 & 97.8 & 97.6 \\
        $\nu\nu ZZ$  & 1557.9 & 121.2 & 36.7 & 36.6 \\
        $\nu W^\pm h$  & 560.7 & 26.2 & 17.2 & 17.2 \\
        $\nu W^\pm Z$  & 492.3 & 12.3 & 9.3 & 9.3 \\
    \hline
        Bkg. & - & - & - & 160.7 \\
    \hline
    \end{tabular}
\end{table}
\begin{table}
    \centering
    \caption{Cut flow for the SM $\nu_e\bar{\nu}_\mu h h$  di-Higgs signal and backgrounds cross sections at \emc with $E_e=100\gev$, $E_\mu=2.5\tev$ (left) and $E_e=170\gev$, $E_\mu=6\tev$ (right).}
    \label{tab:cutflow2}
    \begin{tabular}{|c|c|c|c|}
    \hline
        $\sigma$ [ab] & pre-cuts & $m_{bb}$ cut & HHCUT\\
    \hline
        $\nu\nu hh(\to bbbb)$  & 12.2 & 6.8 & 5.5\\
    \hline              
        $\nu\nu hZ$  & 49.8 & 9.3 & 5.2\\
        $\nu\nu ZZ$  & 66.1 & 8.3 & 2.8\\
        $\nu W^\pm h$  & 36.8 & 2.8 & 1.4\\
        $\nu W^\pm Z$  & 40.2 & 2.2 & 0.9\\
    \hline
        Bkg. & - & - & 10.4\\
    \hline
    \end{tabular}\hspace{1cm}
    \begin{tabular}{|c|c|c|c|}
    \hline
        $\sigma$ [ab] & pre-cuts & $m_{bb}$ cut & HHCUT\\
    \hline
        $\nu\nu hh(\to bbbb)$  & 69.9 & 37.6 & 31.5\\
    \hline              
        $\nu\nu hZ$  & 205.5 & 33.1 & 19.3\\
        $\nu\nu ZZ$  & 298.1 & 30.6 & 10.4\\
        $\nu W^\pm h$  & 143.6 & 9.9 & 6.6\\
        $\nu W^\pm Z$  & 139.6 & 6.5 & 2.8\\
    \hline
        Bkg. & - & - & 39.1\\
    \hline
    \end{tabular}
\end{table}

\subsection{Asymmetric Colliders}
The electron-muon ($e\mu$) collider, initially proposed in Refs.~\cite{Hou:1996ys, Choi:1997bm, Barger:1997dv}, has drawn attentions recently~\cite{Lu:2020dkx}. As a hybrid option, it is often considered to benefit from both the clean environment of electron colliders and the high beam energy of muon colliders. 
In this work, we use two benchmark colliding energies to illustrate the potential of the \emc for Higgs coupling measurements:
\begin{itemize}
\item A conventional 1 TeV setup with $E_e=100\gev$ and $E_\mu=3\tev$,
\item A maximal colliding energy of 2 TeV with $E_e=170\gev$ and $E_\mu=6\tev$.
\end{itemize}
The di-Higgs signal at the \emc is $4b+$\met, similar to that at the \eec, and the background processes are also comparable with the \eec and the \mmc. However, due to non-zero lepton numbers in the initial state, the \emc is free from the SM $s$-channel backgrounds. Besides, there are no backgrounds from $ZZ/Zh$ production at the \emc, and no need for a recoil mass cut. The collider simulation is similar to that of the \eec. The calorimeter coverage range of the detector is set as $-4<\eta<6$ to simulate the energy asymmetric behavior of the \emc, where $\eta$ is the rapidity of the final state objects. The following pre-selection cuts (pre-cuts) are applied:
\begin{align}
&n^\ell(p_T>10\gev)=0,~\met>10\gev,\nonumber \\
&p_T^{jet}>15\gev, -3.0<\eta^{jet}<5.0,~\Delta R^{mn} >0.5.
\end{align}

We then apply the same invariant mass cut and HHCUT as used for the symmetric colliders, with the cut flows displayed in Table~\ref{tab:cutflow2}. 
With an integrated luminosity of $3\ab^{-1}$, the signal strength of the total cross section for di-Higgs production, $\mu_{hh}=\sigma_{hh}/\sigma_{hh}^{\rm SM}$, can be constrained to $0.38<\mu_{hh}<1.78$ at the $1\tev$ \emc and $0.75<\mu_{hh}<1.25$ at the $2\tev$ \emc at 95\% confidence level. 

\section{Analysis}\label{sec:analysis}

\begin{figure}[h!]
    \centering
    \includegraphics[scale=0.35]{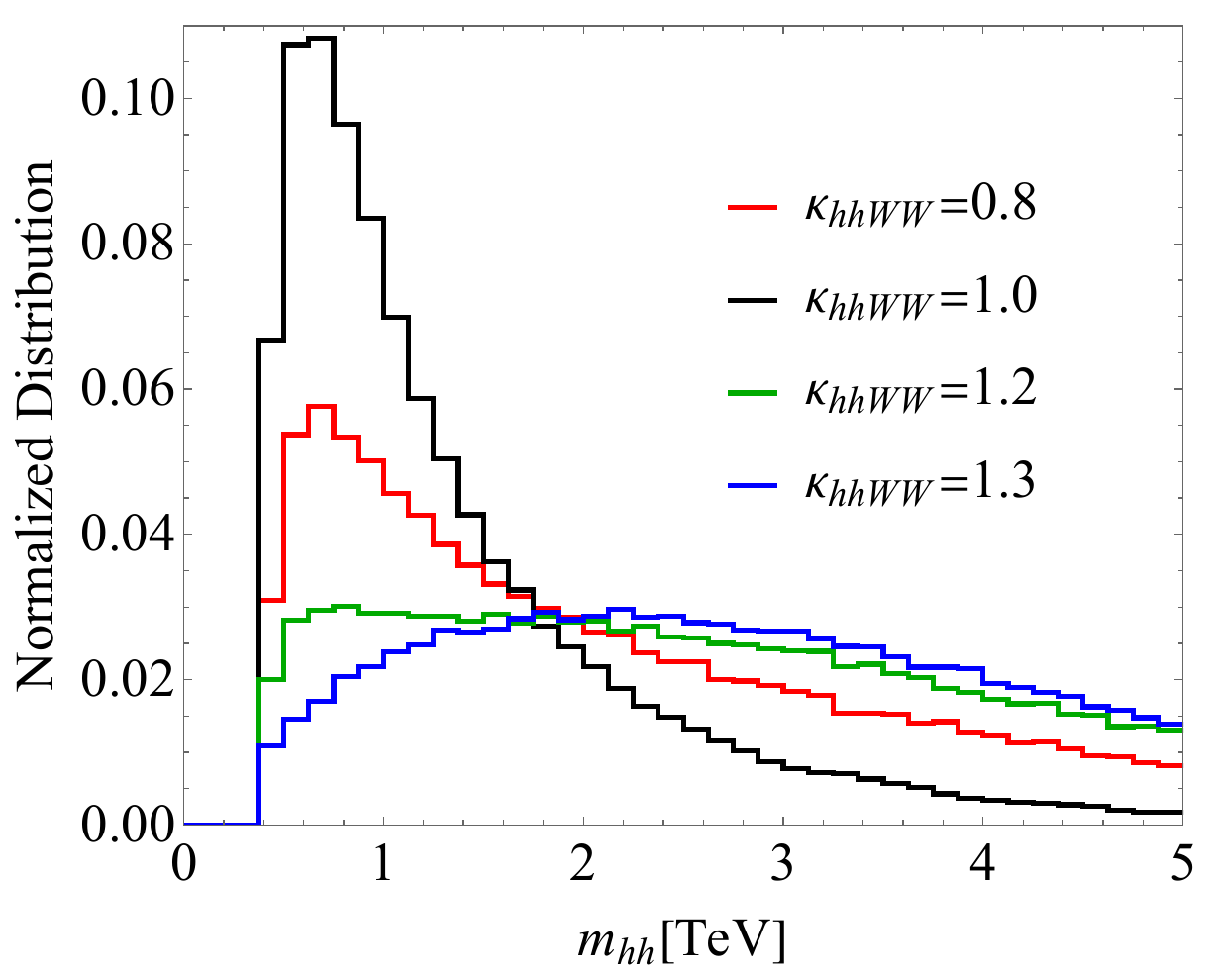}
    \caption{Normalized $m_{hh}$ distribution at 10 TeV \mmc for different \khhww, assuming \kh and \khww to be the SM value.  
     }
    \label{fig:kappabins}
\end{figure}

Equipped with the simulation results, we explore the discovery potential of lepton colliders on Higgs-gauge coupling, equivalently, \khhww and \khww. As stated before, to measure \khhww from the di-Higgs production, the \khww must first be measured from the single Higgs production.  However, with a fixed value of \khww, the total di-Higgs cross section is a quadratic function of \khhww; see Eq.~\eqref{eq:sigmallvvhh1}. Therefore, the measurement of \khhww may exhibit degeneracy. 
The degeneracy can be resolved by analyzing the $m_{hh}$ distribution.  As shown in Fig.~\ref{fig:kappabins}, the $m_{hh}$ distribution for different \khhww are significantly different, which is a consequence of the unitarity violation when $\khhww\neq 1$; see Appendix~\ref{app:unitarity} for details.

\begin{figure}
    \centering
    \includegraphics[scale=0.45]{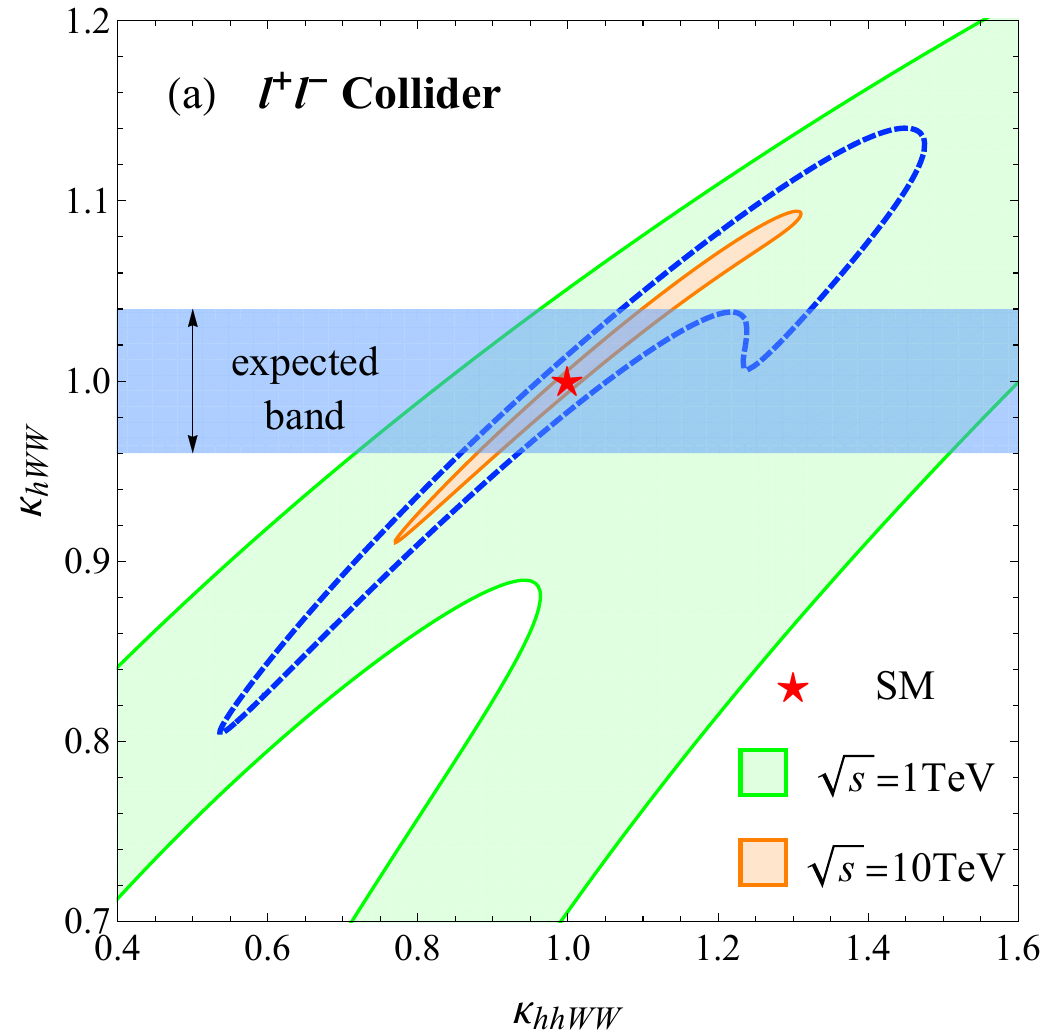}
    \includegraphics[scale=0.45]{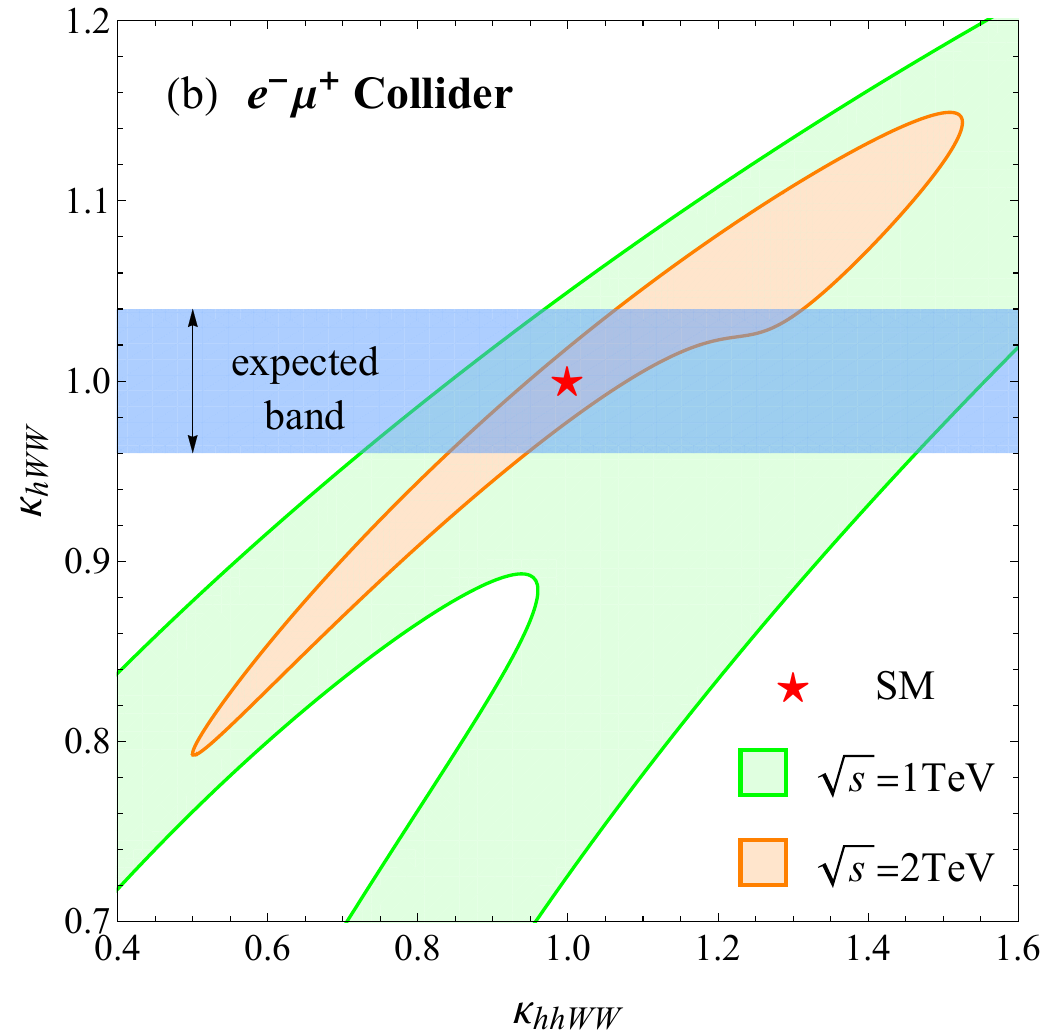}
    \caption{Parameter space of \khww and \khhww that is consistent with SM (\khww=\khhww=1) at 95\% C.L. The left panel gives the results from $1 \tev$ \  \eec and $10 \tev$\  \mmc, the right panel is the case of \emc with $\sqrts = 1, 2 \tev$. The dashed contour in left panel stands for the discovery capability at the 10 TeV \mmc.}
    \label{fig:kappaps}
\end{figure}

In the following, we perform a binned analysis which utilizes the information of $m_{hh}$ distribution.  The likelihood function for a Poisson distribution is given by
\begin{equation}\label{eq:likelihood}
    L(\vec \kappa)=\frac{\left(n_{\rm th}(\vec \kappa)\right)^{n_{\rm obs}} e^{-n_{\rm th}(\vec \kappa)}}{ n_{\rm obs}!},
\end{equation}
where $\vec \kappa=(\khww,\khhww)$ stands for the parameters in the theory, $n_{\rm th}(\vec \kappa)$ is the events number predicted by the theory and $n_{\rm obs}$ is the observed events number. 
For the di-Higgs boson production, the events number from theory prediction is $n_{\rm th}(\vec \kappa)=\left[\sigma_{hh}(\vec \kappa)+\sigma_{\rm bkg}\right]\mathcal{L}$ with an integrated luminosity of $\mathcal{L}$, where $\sigma_{hh}$ is the di-Higgs production cross section after applying selection cuts and $\sigma_{\rm bkg}$ is corresponding backgrounds. The logarithm of likelihood function can be defined as 
\begin{align} \label{EQ:llh}
    -2 \ln \lambda(\vec\kappa_1|\vec\kappa_0)=-2 \ln \frac{L(\vec\kappa_1)}{L(\vec\kappa_0)},
\end{align}
which depicts the hypothesis $\vec\kappa_1$ is excluded versus $\vec\kappa_0$ with $\sqrt{ -2 \ln \lambda(\vec\kappa_1|\vec\kappa_0) }\sigma$ confidence level.  For binned events, the likelihood function in Eq.~\eqref{eq:likelihood} is modified as
\begin{equation}
    L(\vec \kappa)=\prod_i^N\frac{\left(n^i_{\rm th}(\vec \kappa)\right)^{n^i_{\rm obs}} e^{-n^i_{\rm th}(\vec \kappa)}}{ n^i_{\rm obs}!},
\end{equation}
where $n_{\rm th}^i(\vec \kappa)$ and $n_{\rm obs}^i$ are the theory predicted and observed event numbers in the $i^{\rm th}$ bin.

With the likelihood presented above, we discern the potential for measuring Higgs-gauge-boson couplings.  The region of $-2\ln \lambda(\vec{\kappa}|\vec\kappa_{\rm SM})\leq 4$ is depicted in Fig.~\ref{fig:kappaps}.
Figure~\ref{fig:kappaps}(a) displays results from the \eec and the \mmc. The green region represents the $\khhww-\khww$ parameter space that is consistent with the SM at $95\%$ confidence level (C.L.) at the 1 TeV \eec, while the orange region denotes the same for the 10 TeV \mmc. The red star marks the SM prediction where \khww=\khhww=1. The dashed contour in Fig.~\ref{fig:kappaps}(a) stands for the $5 \sigma$ discovery capability at the 10 TeV \mmc. The blue band denotes the 95\% C.L. expected constraint on \khww from the future electron-proton collider, namely $0.96 < \khww < 1.04$ \cite{Sharma:2022epc}, while the current limit on \khww at the 13 TeV LHC is $0.93 < \khww < 1.22 $~\cite{CMS:2018uag}. 
Figure~\ref{fig:kappaps}(b) presents results from asymmetric colliders, specifically the \emc with $\sqrts=1$ TeV (the green region) and $\sqrts=2$ TeV (the orange region).

 From these figures, it is evident that the measurements become more accurate with increasing collision energy due to the rise in event numbers with $\sqrts$, regardless of symmetric or asymmetric colliders.  For example, in the \eec and \emc with $\sqrts=1$ TeV, the signal and background event numbers are similar, as seen in Table~\ref{tab:cutflow2}. Consequently, it is expected that the potentials are comparable, as shown by the green range in Fig.~\ref{fig:kappaps}(a) and (b).

As shown in Fig.~\ref{fig:kappaps}, with the expected limit on \khww, we obtained the expected discovery sensitivity on \khhww as follows:
\begin{itemize}
    \item $5\sigma$ discovery region \khhww $< 0.86$ or \khhww$>1.32$ at the 10 TeV \mmc,
    \item $2\sigma$ exclusion region \khhww $<0.88$ or \khhww $>1.14$ at the 10 TeV \mmc,
    \item $2\sigma$ exclusion region \khhww $<0.72$ or \khhww $>1.70$ at the 1 TeV \eec,
     \item $2\sigma$ exclusion region \khhww $<0.72$ or \khhww $>1.66$ at the 1 TeV \emc,
     \item $2\sigma$ exclusion region \khhww $<0.84$ or \khhww $>1.32$ at the 2 TeV \emc.
\end{itemize}

\section{Conclusion}\label{sec:conclusion}

The quartic Higgs-gauge-boson coupling $g_{hhVV}$ is essential to the electroweak symmetry breaking mechanism and gauge boson mass generation, and it can be measured in the VBF-type di-Higgs production. However, at hadron colliders, the process is typically overshadowed by the gluon-fusion di-Higgs production, making the extraction of $g_{hhVV}$ challenging. In contrast, lepton colliders offer an ideal platform for probing the VBF-type di-Higgs production.

In the work we explored the potential of lepton collider on the quartic Higgs-gauge-boson coupling, or equivalently, its ratio to SM prediction \khhww. Detailed collider simulations including parton showering effects were performed at the symmetric \eec, \mmc and the asymmetric \emc. We found that the collider's potential to constrain \khhww primarily depends on the collision energy. With an integrated luminosity of 3 ab$^{-1}$, the $2\sigma$ exclusion region is \khhww $<0.72$ or \khhww $>1.70$ and \khhww $<0.72$ or \khhww $>1.66$ at the 1 TeV \eec and \emc ($E_e=100\gev,E_\mu=3\tev$), respectively. At the 2 TeV \emc ($E_e=170\gev,E_\mu=6\tev$), the $2\sigma$ exclusion region is \khhww $<0.84$ or \khhww $>1.32$, and it is \khhww $<0.88$ or \khhww $>1.14$ at 10 TeV \mmc. In addition we obtain the $5\sigma$ discovery region of \khhww $<0.86$ or \khhww $>1.32$ at the 10 TeV \mmc.

\vspace{10pt}
{\bf{Note added:}} Near the completion of this work, Ref.~\cite{Davila:2023fkk} appeared with related content.

\begin{acknowledgements}
The work of QHC, KC and XRW is partly supported by the National Science Foundation of China under Grant No. 11725520 and No. 12235001. The work of YDL is partly supported by the National Science Foundation of China under Grant No. 12075257.
\end{acknowledgements}
\appendix

\section{Perturbative Unitarity}\label{app:unitarity}
\begin{figure}
    \centering
    \includegraphics[width=\linewidth]{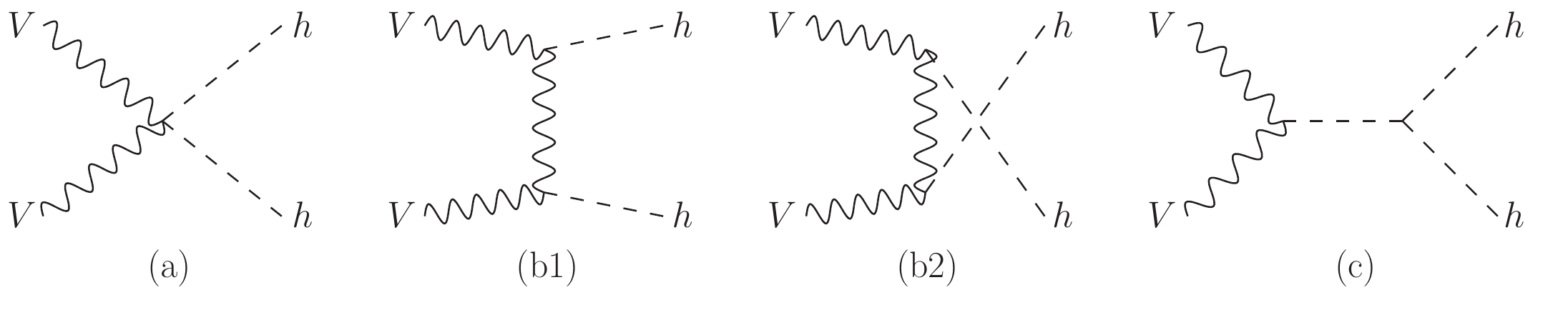}
    \caption{Feynman diagrams representing $VV\to hh$ ($V=W^\pm/Z$) processes. These correspond to the factorized subprocess shown in Fig.~\ref{fig:ll2HH}.}
    \label{fig:VV2HH}
\end{figure}

\begin{figure}
    \centering
    \includegraphics[width=.6\linewidth]{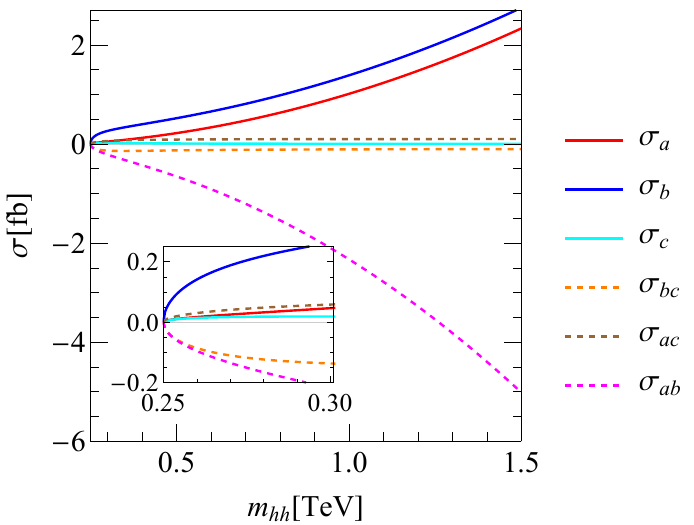}
    \caption{The energy dependence of each Feynman diagram contribution of the subprocess $W^+_L W^-_L\to hh$. The solid line denotes the contribution from Feynman diagram (a), (b) and (c), respectively; and the dashed line the interference contribution.     }
    \label{fig:wlwl2hh}
\end{figure}

The nature of the VBF process can be ascertained by the factorized subprocess of $W^-W^+\to hh$. This is executed through the Effective-$W$ approximation (EWA) method~\cite{Dawson:1984gx} and the cross section is expressed as the equation below:
\begin{equation}\label{eq:EWAint}
    \sigma(\ell^-\ell^+\to \nu_\ell \bar{\nu}_\ell hh)=\sum_{i,j}\int dx_1 dx_2 P^i_{W^-/\ell^-}(x_1) P^j_{W^+/\ell^+}(x_2) ~\sigma(W^-_i W^+_j\to hh).
\end{equation}
Here $P^i_{W^-/\ell^-}(x_1)$ denotes the gauge boson $W^-$'s probability distribution with polarization $i$ in the electron; $x_1=(E_{\ell^-}-E_{\nu_\ell})/E_{\ell^-}$ is the energy fraction of $\ell^-$ carried by $W^-$.  The splitting functions for the collinear radiation of real $W$-boson at leading order are~\cite{Barger:1990py}.
\begin{align}\label{eq:EWAPDF}
& P_{W / \ell}^{+}(x)=\frac{g^2}{32 \pi^2 x} \ln \left(\frac{4x^2 E_{\ell}^2}{M_W^2}\right), \nonumber \\
& P_{W / \ell}^{-}(x)=\frac{g^2}{32 \pi^2 x}(1-x)^2 \ln \left(\frac{4x^2 E_{\ell}^2}{M_W^2}\right), \nonumber \\
& P_{W / \ell}^L(x)=\frac{g^2}{16 \pi^2 x}(1-x).
\end{align}
The matrix element of the subprocess $W^\mu(q_1) W^\nu(q_2)\to h(k_1)h(k_2)$ is
\begin{align} \label{eq:MEvv2hh}
    M^{\mu\nu}=&\left[
    \frac{3m_h^2 g_{hWW}}{(\hat s-m_h^2)v} + 2g_{hhWW} + \left(\frac{g_{hWW}^2}{\hat{t}-m_W^2}+\frac{g_{hWW}^2}{\hat{u}-m_W^2}\right)
    \right]g^{\mu\nu} - \nonumber\\
    &\frac{g_{hWW}^2}{m_W^2}
    \left[  \frac{(q_1-k_1)^\mu(q_1-k_1)^\nu }{\hat{t}-m_W^2}+\frac{(q_1-k_2)^\mu(q_1-k_2)^\nu }{\hat{u}-m_W^2}  \right],
\end{align}
where $\hat{s}$ ($\hat{t}$, $\hat{u}$) are Mandelstam variables of the subprocess, and $g_{hWW}$ and $g_{hhWW}$ are gauge couplings of the Higgs boson with $W^\pm$ boson. The SM presents the couplings in the following:
\begin{align}
  g_{hhWW}^{\rm SM}=\frac{m_W^2}{v^2},\quad g_{hWW}^{\rm SM}=2\frac{m_W^2}{v},
\end{align}
where $v=246\gev$ refers to the electroweak symmetry breaking scale. The couplings are parametrized as $g_{hWW}=\khww g_{hWW}^{\rm SM}$ and $g_{hhWW}=\khhww g_{hhWW}^{\rm SM}$. The cross section of the subprocess $W^+W^-\to hh$ can then be parametrized as follows:
\begin{align}\label{eq:WWhh}
    \sigma(W^+W^-\to hh) =& \khhww^2 \sigma_a + \khww^4 \sigma_b + \khww^2 \sigma_c +\khhww\khww^2\sigma_{ab} +\nonumber \\
    &\khhww\khww\sigma_{ac}+\khww^3\sigma_{bc},
\end{align}
where $\sigma_a$, $\sigma_b$, and $\sigma_c$ parametrize the contribution from Feynman diagrams (a), (b), and (c) depicted in Fig.~\ref{fig:ll2HH}, whereas $\sigma_{ab}$, $\sigma_{ac}$, and $\sigma_{bc}$ are the corresponding interferential contributions. The cross section of the complete process, i.e., $\ell^-\ell^+\to \nu_\ell \bar{\nu}_\ell hh$, can be parametrized similarly. The dependence of the cross section of $W^+W^-\to hh$ on the couplings \khww and \khhww can be inferred from the parameters in Eq.~\eqref{eq:WWhh}. As shown in Fig.~\ref{fig:wlwl2hh}, the contributions from Feynman diagrams (a) and (b) in Fig.~\ref{fig:VV2HH} and their corresponding interference components dominate the cross section. 

In high-energy limits, the amplitude can be depicted by:
\begin{align} \label{eq:ampLL}
\mathcal{M}_{LL}=M^{\mu\nu} \epsilon^{\mu}_L(q_1) \epsilon^{\nu}_L(q_2)=\frac{\hat{s}}{v^2}\big(\kappa_{hhWW}-\kappa_{hWW}^2\big) +\mathcal{O}(\hat{s}^0),
\end{align}
for the longitudinally polarized vector bosons. This notion holds true as the longitudinal polarization vector $\epsilon^\mu_L(q)$ approximates to $q^\mu/m_W$. This expression reveals that high energy collision regions primarily receive contributions from Feynman diagrams (a) and (b), a finding that is consistent with the above numeric result. Furthermore, the violation of the unitarity bound is inevitable unless the sum rule detailed below is fulfilled:
\begin{align} \label{eq:sumrule}
\kappa_{hhWW} - \kappa_{hWW}^2=0.
\end{align}

\begin{figure}
    \centering
    \includegraphics[scale=0.55]{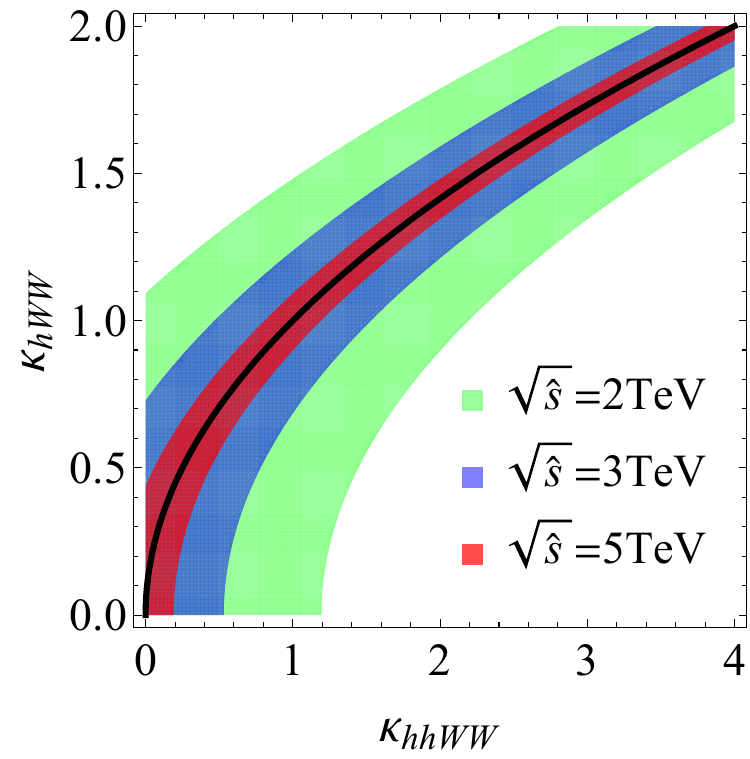}
    \caption{Parameter space of \khhww and \khww that satisfies unitarity.} 
    \label{fig:kappaPS}
\end{figure}

Perturbative unitarity requires the partial wave amplitude $a_l$~\cite{Lee:1977yc,Lee:1977eg} to satisfy
\begin{align}
|\operatorname{Re}{a_l}|\leq 1/2,
\end{align}
with the definition of
\begin{align}
a_l =\frac{1}{32\pi} \int_{-1}^{1}{\mathcal{M}}(\hat{s},\hat{t}) {\text{P}}_l(\cos\theta) \operatorname{d}\cos\theta,
\end{align}
and $\mathcal{M}(\hat{s}, \hat{t})$ is amplitude of the process $W^-W^+\to hh$ given in Eq.~\eqref{eq:ampLL}. Consequently, the unitarity bound constrains the parameters \khww and \khhww for a given collision energy $\sqrt{\hat{s}}$ of the subprocess; see Fig.~\ref{fig:kappaPS}. 
The colored regions in the figure represent the unitarity-consistent areas for various collision energies, while the solid black line indicates the sum rule in Eq.~\eqref{eq:sumrule}. The figure demonstrates that, as $\sqrt{\hat{s}}=m_{hh}$ increases, deviations from the sum rule must become less severe to maintain the theory consistence.

\bibliographystyle{apsrev4-1}
\bibliography{ref}
\end{document}